\documentclass[runningheads]{llncs}

\usepackage{amssymb}
\usepackage{graphicx}

\usepackage{url}
\urldef{\mailsa}\path|{kulakowski,3natkani}@novell.ftj.agh.edu.pl|
\newcommand{\keywords}[1]{\par\addvspace\baselineskip
\noindent\keywordname\enspace\ignorespaces#1}

\begin{document}

\mainmatter  

\title{When the spatial networks split?}


%
%
\author{Joanna Natkaniec and Krzysztof Ku{\l}akowski%
}
%

\institute{Faculty of Physics and Applied Computer Science, AGH--UST,\\
 al. Mickiewicza 30, PL-30059 Krak\'ow, Euroland\\
\mailsa\\
\url{http://www.zis.agh.edu.pl/}}

%
%

\toctitle{Lecture Notes in Computer Science}
\tocauthor{Authors' Instructions}
\maketitle

\begin{abstract}
We consider a three dimensional spatial network, where $N$ nodes are randomly distributed within a cube $L\times L\times L$. Each two nodes 
are connected if their mutual distance does not excess a given cutoff $a$. We analyse numerically the probability 
distribution of the critical density $\rho_c=N(a_c/L)^3$, where one or more nodes become separated; $\rho_c$ is found to increase 
with $N$ as $N^{0.105}$, where $N$ is between 20 and 300. The results can be useful for a design of protocols to control sets of wearable sensors.
\end{abstract}

\section{Introduction}

Recent interest in abstract networks is at least partially due to the fact that they are not bound by geometry. However, 
in many applications the networks are embedded in a metric space; then we speak on spatial networks \cite{prov}, geographical
networks \cite{huang}, ad-hoc networks \cite{stepan} or random geometric graphs \cite{dall}. If the connections between nodes
are determined by their mutual distance, this embedding appears to be important for the properties of the system. The list
of examples of spatial networks includes the Internet, the electriticity power grid, transportation and communication networks 
and neuronal networks.

We consider a three dimensional spatial network, where $N$ nodes are randomly distributed within a cube $L\times L\times L$. 
Each two nodes are connected if their mutual distance does not excess a given cutoff $a$ \cite{prov}. In the case of uniform
density of nodes the small-world property is absent, because the average shortest path increases linearly with the system size $L$. 
Here the small world effect could appear in the case of unlimited dimensionality $D$ of the network; once $D$ is fixed, the 
effect disappears. In literature, most papers are devoted to percolation problems; references can be found in \cite{prov,dall}.
The percolation threshold can be identified with the critical density, where the size of the largest connected component becomes
of the order of the number of all nodes. Here we search for another critical density, where the size of the largest connected 
component becomes equal to the number of all nodes. In other words, we investigate the critical spatial density where at least
one node is unconnected. 

In our opinion, the problem can be relevant for control sets of wearable sensors \cite{luk}. To give an example, a group of divers 
wants to keep contact, operating in dark water. Their equipment secures communication between two divers only if the distance 
between them is short enough. In this example, it is crucial to maintain communication with all divers; no one can be lost. 
Having $N$ divers, how large volume of water can be safely penetrated? 

The problem has its physical counterpart; we can ask about the largest fluctuations of the density of an ideal gas. The probability 
distribution of this density is usually assumed to be Gaussian \cite{reichl}. Then we should ask about the probability that the minimal density 
in a gas of $N$ particles is not lower than some critical value proportional to $a^{-3}$. This question belongs to the statistics of 
extremes \cite{gumbel,coles}. However, the derivation of the Gaussian distribution itself relies on the assumption that different areas
in the gas are statistically independent; in real systems this is not true. Then it makes sense to investigate the problem numerically.

In the next section we describe the details of our calculations and the results. Short discussion in the last section closes the text.

\section{Calculations and results}

A set of $N$ points are randomly distributed with uniform probability in a cube $L\times L\times L$. We set $L=4$ and we 
vary $a$ and $N$; then the density is $\rho=N(a/L)^3$. A link is set between each two nodes, if their mutual distance is less than $a$. 
The simplest method is to generate the positions of the nodes and to vary the radius $a$; for each $a$, the connectivity matrix is 
created and investigated. 

\begin{figure} 
\vspace{0.3cm} 
{\par\centering \resizebox*{8cm}{6cm}{\rotatebox{0}{\includegraphics{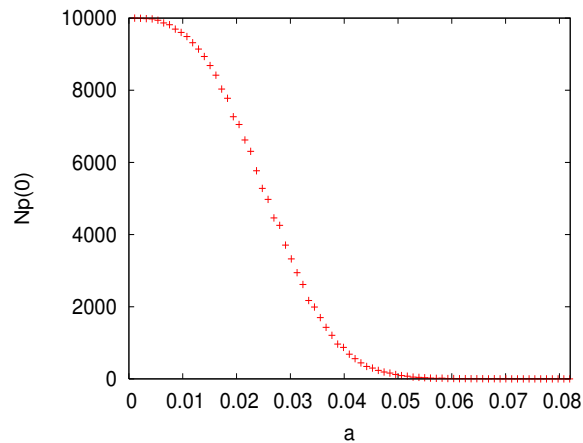}}} \par} 
\vspace{0.3cm} 
\caption{The number of unconnected nodes against the radius $a$ for $N=10^4$.}
\label{fig-1}
\end{figure}

\begin{figure} 
\vspace{0.3cm} 
{\par\centering \resizebox*{8cm}{6cm}{\rotatebox{0}{\includegraphics{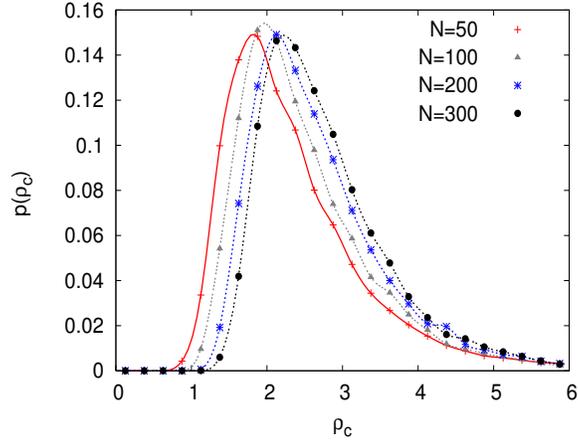}}} \par} 
\vspace{0.3cm} 
\caption{The probability distribution of the critical density $\rho_c$ for various $N$.}
\label{fig-2}
\end{figure}

\begin{figure} 
\vspace{0.3cm} 
{\par\centering \resizebox*{8cm}{6cm}{\rotatebox{0}{\includegraphics{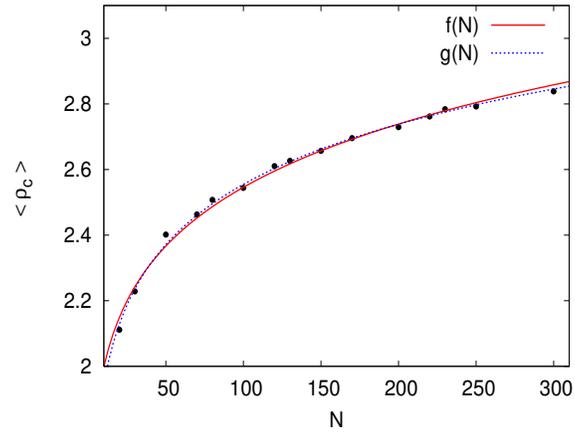}}} \par} 
\vspace{0.3cm} 
\caption{The averaged critical density $\rho_c$ against the system size $N$. The results can be fitted equally well with the functions 
$f(N)=1.56755 \times N^{0.105309}$ and $g(N)=0.264698 \times \ln(N)+1.33571$. }
\label{fig-3}
\end{figure}

\begin{figure} 
\vspace{0.3cm} 
{\par\centering \resizebox*{8cm}{6cm}{\rotatebox{0}{\includegraphics{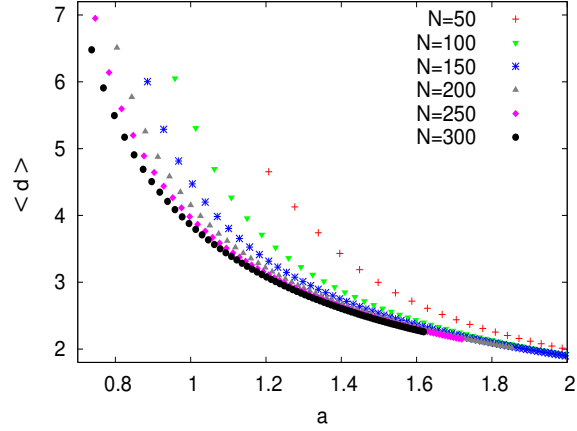}}} \par} 
\vspace{0.3cm} 
\caption{The average value of the network diameter $d$ against the radius $a$ for various $N$.}
\label{fig-4}
\end{figure}

\begin{figure} 
\vspace{0.3cm} 
{\par\centering \resizebox*{8cm}{6cm}{\rotatebox{0}{\includegraphics{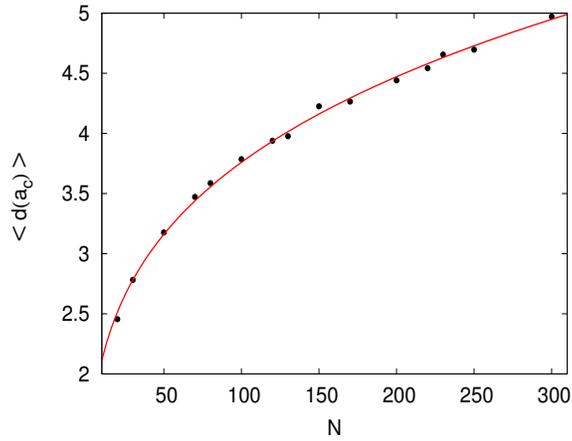}}} \par} 
\vspace{0.3cm} 
\caption{The mean value of the network diameter $d$ at the critical radius $a_c$ as dependent on the system size $N$. 
The results can be fitted with $f(N)=1.18834 \times N^{0.250106}$. This curve does not depend on the model value of $L$.}
\label{fig-5}
\end{figure}

\begin{figure} 
\vspace{0.3cm} 
{\par\centering \resizebox*{8cm}{6cm}{\rotatebox{0}{\includegraphics{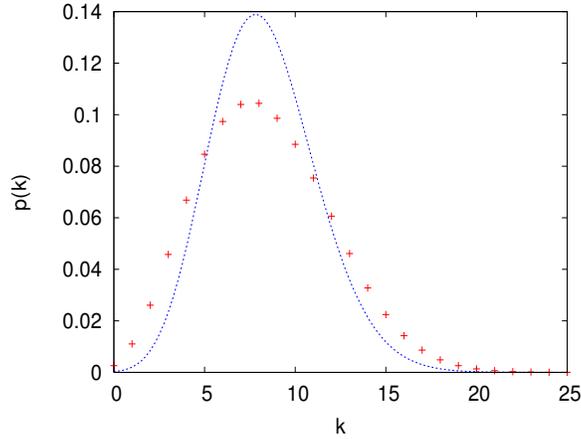}}} \par} 
\vspace{0.3cm} 
\caption{The degree distribution averaged over $10^4$ networks of $N=170$ nodes at $a_c=1.005$ (crosses), compared with the Poisson distribution 
with the same mean degree $<k>=8.322$. }
\label{fig-6}
\end{figure}

We are interested in the critical density $\rho_c$, where some nodes become unconnected with the others. The simplest way is to calculate
the percentage $p$ of isolated nodes per $N$; if there is a phase transition, $p$ could play a role of the order parameter. The problem 
is that in this way, splittings of the network into larger pieces is disregarded; the advantage is that the code works quickly and larger 
networks can be investigated. In Fig. 1 we show $Np$ as dependent on $a$ for $N=10^4$; as we see, the variation is not sharp. Therefore 
we cannot decide if there is a phase transition or just a crossover.

Other results are obtained for smaller lattices, but in each case the algorithm detects the splitting of the whole network into pieces
of any size. In Fig. 2 we present the probability distribution of the critical density $\rho_c$ for selected sizes of the system.
For each $N$, these results were obtained from $10^4$ randomly generated networks. In Fig. 3 we show the mean value of $\rho_c$, as 
it increases with the system size $N$. This dependence appears to be very slow; it can be fitted as proportional to $N^{0.105}$
or, alternatively, as $1.336+0.265\times \ln(N)$. 

We calculate also the network diameter $d$; this is the mean shortest path between nodes, calculated as the number of links between
them. Obviously, $d$ decreases with $a$, and it becomes infinite when the network is disconnected. The calculations are done with the
Floyd algorithm \cite{reftofloyd}, for $a>a_c$. The results are shown in Fig. 4. Fig. 5 reproduces the values of $d$ at the threshold,
where the network splits. 

In Fig. 6 we show the degree distribution of the network for $N=170$ and $a=a_c=1.005$. As we see, the distribution differs from 
the Poisson distribution with the same mean degree. This difference may be due to the correlations between numbers of nodes in different
spheres. 

\section{Discussion}

It is obvious that the probability that at some point the density will be lower than the critical value increases with the system size.
This increase is compensated by the decrease of the critical density and, subsequently, an increase of the critical cutoff with $N$.
The question is how this cutoff increases. The result shown in Fig. 3 indicate, that this increase is rather slow. Our numerical
method does not allow us to differ between the power law with a small exponent and the logarithmic law. 

The data of the mean free path $d$ as dependent on $a$ can be used to design protocols to communication between sensors. An example of such 
a protocol could be that a signal 'zero' detected by one sensor is sent to its neighbors which after a time $\tau$ should reproduce it once, 
adding plus one to the content. The number when the last sensor gets the signal is just the shortest path between this sensor and the one which 
initialized the series. At the threshold density, the communication is partially broken; high value of $d$ near the threshold should activate
the message 'go back to the others' at these sensors which get the high value of the signal. Here, the data shown in Fig. 5 can be useful. How to 
design routing protocols in ad hoc networks is a separate branch of computer science \cite{pro}.

In fact, our Monte Carlo simulations can be seen as an attempt to sample the phase space of the system. The probability that the contact 
is broken can be interpreted dynamically as the percentage of time when the communication is incomplete. If the trajectory wanders randomly, 
all nodes can happen to be connected again.

\bigskip

\section*{Acknowledgements} The authors are grateful to Sergei N. Dorogovtsev and Paul Lukowicz for helpful suggestions.

\end{document}